\documentclass[aps,prl,reprint,superscriptaddress,longbibliography,floatfix,nofootinbib]{revtex4-2}
\usepackage{array}[=2016-10-06]
\usepackage{amsmath,amssymb,bm,booktabs,color,graphicx}
\usepackage[colorlinks=true,linkcolor=blue,citecolor=blue,urlcolor=blue]{hyperref}
\hypersetup{pdftitle={Can Nuclear Recoils and Solar Neutrinos Resolve the Composition of Singlino-Higgsino Dark Matter?},pdfauthor={Jingwei Lian and Jin Min Yang}}
\newcommand{\GeV}{\mathrm{GeV}}
\newcommand{\keV}{\mathrm{keV}}

\newcommand{\sectionlead}[1]{\par\noindent\hspace*{\parindent}\textit{#1}---}
\newcommand{\subsectionlead}[1]{\par\noindent\hspace*{\parindent}\textit{#1}.~}

\usepackage{etoolbox}
\hypersetup{hypertexnames=false}
\makeatletter
\let\auto@bib@innerbib\@empty
\let\write@bibliographystyle\relax
\let\combined@smfootnotesize\footnotesize
\let\combined@author\author
\let\combined@affiliation\affiliation
\let\combined@email\email
\let\combined@thanks\thanks
\let\combined@and\and
\edef\combined@topnumber{\the\c@topnumber}
\edef\combined@bottomnumber{\the\c@bottomnumber}
\edef\combined@dbltopnumber{\the\c@dbltopnumber}
\let\combined@topfraction\topfraction
\let\combined@bottomfraction\bottomfraction
\let\combined@textfraction\textfraction
\let\combined@dbltopfraction\dbltopfraction
\makeatother

\renewcommand{\footnotesize}{\fontencoding{OT1}\fontfamily{cmr}\fontsize{8}{9.5}\selectfont}
\renewcommand{\topfraction}{0.95}
\renewcommand{\bottomfraction}{0.85}
\renewcommand{\textfraction}{0.05}
\renewcommand{\dbltopfraction}{0.95}
\begin{document}
\title{Can Nuclear Recoils and Solar Neutrinos Resolve the Composition of Singlino-Higgsino Dark Matter?}
\author{Jingwei Lian}
\email{lianjw@hist.edu.cn}
\affiliation{Henan Institute of Science and Technology, Xinxiang 453003, P. R. China}
\affiliation{Centre for Theoretical Physics, Henan Normal University, Xinxiang 453007, P. R. China}
\author{Jin Min Yang}
\email{jmyang@itp.ac.cn}
\affiliation{Centre for Theoretical Physics, Henan Normal University, Xinxiang 453007, P. R. China}
\affiliation{Institute of Theoretical Physics, Chinese Academy of Sciences, Beijing 100190, P. R. China}
\begin{abstract}
A nuclear recoil measures dark matter (DM) scattering on nucleons, but usually cannot reveal the nature of the halo DM. An interchange of the ground and excited states of the DM can preserve an inelastic recoil spectrum while changing ground-state annihilation by several orders of magnitude. Solar neutrinos can then provide information that nuclear recoils leave unresolved. Motivated by the high-energy LUX-ZEPLIN candidate, we investigate a possible way to resolve the composition of singlino--Higgsino DM in the next-to-minimal supersymmetric standard model (NMSSM) by using nuclear recoils and solar neutrinos. Our analysis identifies solar-compatible singlino ground states and Higgsino ground states with weak inelastic transitions, whereas Higgsino-pair inelastic LZ fits predict excessive solar signals. An elastic Higgsino scattering alternative can suppress solar capture by closing the inelastic transition. Recoil spectra, solar evolution and target complementarity can therefore probe distinct aspects of the singlino--Higgsino DM composition, with residual degeneracies set by annihilation, halo and detector assumptions.
\end{abstract}
\maketitle

\sectionlead{Introduction}
The LUX-ZEPLIN (LZ) Collaboration has reported a nuclear-recoil candidate at $248\pm23\,\mathrm{(stat)}\pm23\,\mathrm{(sys)}\,\keV$ in a $2.84$ tonne-year exposure, with a global background-only tension of $2.6\sigma$~\cite{LZ}. This event has prompted a broad range of dark-matter explanations and complementary tests, encompassing endothermic and exothermic transitions, elastic scattering, absorption and boosted populations~\cite{LZOverview}.

In the minimal supersymmetric standard model (MSSM), nearly degenerate neutral Higgsinos provide the off-diagonal $Z$ interaction. Their thermal abundance selects $m_\chi\simeq1.1\,\mathrm{TeV}$, whereas a few-hundred-keV splitting typically demands gauginos near $10^7\,\GeV$ or cancellation of signed gaugino contributions~\cite{Higgsino,Nagata}. Solar gravity opens otherwise restricted capture channels, making neutrino searches especially relevant~\cite{Solar,IceCube}. The NMSSM dynamically generates the Higgsino mass and adds a singlino that modifies splittings with TeV-scale gauginos~\cite{NMSSM,GNMSSM}. 
Singlino interpretations of the LZ candidate and gamma-ray probes have recently been discussed~\cite{Singlino}, alongside solar-neutrino constraints on elastic bino-like dark matter and inelastic Higgsino pairs~\cite{Roy}. 
The solar-neutrino viability of inelastic singlino dark matter remains to be established. This also raises a more fundamental question: to what extent can LZ, IceCube and future recoil measurements resolve the composition of the ground state and its accessible excitation when different singlino-Higgsino orderings yield similar recoil spectra?

We derive the mixing and splitting relations of three singlino--Higgsino orderings, follow solar capture and annihilation, and examine elastic scattering in the same model domain. An inelastic recoil measures a transition coupling, whereas solar annihilation probes the ground state. This distinction lifts part of the composition degeneracy; an elastic interpretation restores a competing Higgsino explanation.

\sectionlead{Neutralino composition and mass ordering}
We work in the CP-conserving, $R$-parity-conserving $Z_3$-invariant NMSSM, with $\mu=\lambda\langle S\rangle$ and $m_S=2\kappa\mu/\lambda$. The analytic description uses the decoupled-gaugino basis $(\widetilde H_d^0,\widetilde H_u^0,\widetilde S)$.\footnote{The numerical spectrum retains positive bino, wino and gluino soft masses of $3\,\mathrm{TeV}$. Small bino/wino fractions preserve the singlino--Higgsino ordering, but their mass corrections are retained when resolving sub-MeV splittings.} We label the corresponding components of the neutralino mixing matrix by $N_{id}$, $N_{iu}$ and $N_{iS}$, and define
\begin{equation}
\begin{aligned}
 F_{H,i}&=|N_{id}|^2+|N_{iu}|^2,\\
 C_{ij}&=N_{id}N_{jd}^*-N_{iu}N_{ju}^*,~~
 D_1=|N_{1d}|^2-|N_{1u}|^2.
\end{aligned}
\label{eq:definitions}
\end{equation}
In this CP-conserving convention, coherent vector $Z$ transitions connect opposite-sign eigenvalues; physical masses are their absolute values~\cite{Spinors}. For this pair, $f=|C_{12}|^2$ and $\delta_{21}=m_{\chi_2^0}-m_{\chi_1^0}$. Unitarity gives $\sum_j|C_{1j}|^2=F_{H,1}$: the accessible transition alone determines neither the ground-state Higgsino fraction nor $D_1$. We label the three light neutralinos SHH, HSH or HHS according to their predominantly singlino (S) or Higgsino (H) character, ordered in increasing physical mass.

\subsectionlead{SHH: singlino ground state}
A singlino ground state lies below two Higgsinos and mixes mainly with one Higgsino combination. Denoting the mixing entry by $b$ and the same-sign eigenvalue separation by $\Delta_{HS}$, two-state diagonalization gives
\begin{equation}
 b^2=f(1-f)\Delta_{HS}^2,\qquad
 f\simeq F_{H,1}.
\label{eq:mixing}
\end{equation}
The accessible transition connects opposite-sign roots, so $\delta_{21}\ll\Delta_{HS}$ does not require $\lambda v\lesssim\delta_{21}$. For $\tan\beta=10$ and $\Delta_{HS}=3\,\GeV$, $\lambda=10^{-3}$ gives $f\simeq0.002$ and lowers the singlino-like mass by $f\Delta_{HS}\simeq6\,\mathrm{MeV}$. Adjusting $m_S$ to compensate for this mixing correction can yield $\delta_{21}=320\,\keV$. The coupling is moderately small, but the residual degeneracy is tuned rather than symmetry protected (Supplemental Material).

\subsectionlead{HSH: singlino first excitation}
In HSH, a Higgsino ground state precedes a singlino excitation and a second Higgsino. Interchanging the singlino and opposite-sign Higgsino preserves the mixing mechanism, now with $f\simeq F_{H,2}$. The compositions $(F_{H,1},F_{H,2})\simeq(f,1)$ and $(1,f)$ can therefore share a transition strength. Matching absolute eigenvalues fixes their physical gap independently of the distant same-sign mixing partner.

With $F_{H,1}\simeq1$ and $f\ll1$, most neutral-current strength resides in $\chi_1^0\to\chi_3^0$. This channel must remain closed to avoid large solar capture. Electroweak ground-state annihilation and elastic amplitudes independent of $f$ then distinguish HSH from SHH.

\subsectionlead{HHS: Higgsino ground-state pair}
The two lightest states are Higgsino dominated and the singlino-like state is heavier, yielding an approximately unsuppressed coherent transition. Away from resonant singlino mixing, the leading splitting is
\begin{equation}
\delta_{21}\simeq\left|G_0+
\frac{\lambda^2v^2(m_S-\mu\sin2\beta)}{m_S^2-\mu^2}\right|.
\label{eq:HHS}
\end{equation}
Here $G_0=m_Z^2(s_W^2/M_1+c_W^2/M_2)$ is the leading correction from the bino and wino soft masses $M_1,M_2$, and $v=174\,\GeV$. Cancellation between singlino and gaugino terms permits sub-MeV splittings with TeV-scale gauginos; larger gaps instead close the coherent transition and allow an elastic interpretation. 

\begin{table*}[t]
\caption{Representative spectra with Helm coherent and GCN5082 spin responses. BP1 is SHH; $\mathrm{BP2}^{\rm c}/\mathrm{BP3}^{\rm c}$ form the SHH/HSH comparison pair. No evaluated inelastic HHS explanation satisfies the adopted solar limit, so BP4 illustrates the elastic HHS alternative. $N_{\rm low}$ and $N_{\rm high}$ are acceptance-weighted event expectations over $5.4$--$100$ and $100$--$269.9\,\keV$; $N_{\rm SB}$ is the predicted sideband count over $350$--$590\,\keV$, evaluated with unit acceptance. The solar signal of $\mathrm{BP2}^{\rm c}$ is negligibly small.}
\label{tab:bench}
\centering\small\setlength{\tabcolsep}{3pt}
\begin{tabular}{lrrrrrrrr}
\toprule
Point & $m_\chi$ [GeV] & $\delta_{21}$ & $\lambda$ & $N_{\rm low}$ & $N_{\rm high}$ & $N_{\rm SB}$ & $q_{\rm LZ}$ & $\mathcal R_\odot$\\
\midrule
BP1: SHH & 498.618 & $319\,\keV$ & $7.99\times10^{-4}$ & 0.00004 & 0.5277 & 0.03045 & 1.825 & 0.003\\
$\mathrm{BP2}^{\rm c}$: SHH & 738.908 & $228\,\keV$ & $9.23\times10^{-5}$ & 0.21399 & 1.3730 & 0.04384 & 4.055 & $10^{-12}$\\
$\mathrm{BP3}^{\rm c}$: HSH & 737.714 & $225\,\keV$ & $1.34\times10^{-4}$ & 1.48877 & 1.2154 & 0.03841 & 6.669 & 0.762--0.925\\
BP4: HHS & 504.278 & $2.98\,\GeV$ & $4.42\times10^{-1}$ & 3.28319 & 0.6086 & 0.03828 & 9.871 & 0.508--0.777\\
\bottomrule
\end{tabular}
\end{table*}

\sectionlead{Recoil signals and solar evolution}
Endothermic recoils select the high-speed halo tail. We assume ground-state dark matter in a truncated Maxwellian standard halo model (SHM): $\rho_\chi=0.3\,\GeV\,\mathrm{cm}^{-3}$, $v_0=238\,\mathrm{km\,s}^{-1}$ and Galactic escape speed $544\,\mathrm{km\,s}^{-1}$; the Earth and Sun speeds are $254$ and $232\,\mathrm{km\,s}^{-1}$. Coherent $Z$ and elastic spin-independent (SI) scattering use isotope-dependent Helm form factors with weak and scalar charges, respectively. Elastic spin-dependent (SD) scattering uses the GCN5082 xenon response~\cite{PandaXSD,Anand}, retaining signed proton--neutron interference. The LZ comparison includes acceptance, resolution and calibration profiling.

Solar gravity opens additional capture channels through $w^2=u^2+v_{\rm esc,\odot}^2(r)$. At fixed masses and halo, the accepted transition count and its capture rate satisfy
\begin{equation}
 N_Z=fK_{\rm Xe},~~ C_Z=fK_\odot
 ~~\Longrightarrow~~
 C_Z=N_Z\frac{K_\odot}{K_{\rm Xe}}.
\label{eq:capture}
\end{equation}
Lowering mixing therefore cannot suppress this capture contribution at fixed transition signal. Other accessible transitions and elastic scattering add to the total capture.

Solar evolution supplies the composition sensitivity. We follow excitation, de-excitation, decay, elastic cooling, escape and annihilation without imposing a thermal core or number equilibrium on inelastic solutions. Defining $F_\Gamma=2\Gamma_A/C_{\rm tot}$, the neutrino yield in bin $i$ is $N_{\nu,i}=T_\nu(C_{\rm tot}/2)F_\Gamma Y_i$, where $Y_i$ includes branching fractions, source geometry and propagation. The Supplemental Material derives the equilibrium limit and its orbital qualifications.

Away from resonances, weakly mixed SHH has $\langle\sigma v\rangle_{11}\propto f^2$, whereas a Higgsino ground state retains electroweak annihilation as $f\to0$. Coannihilation with Higgsinos, charginos and, where relevant, sleptons can nevertheless maintain the relic abundance~\cite{Coann}. We propagate the model gauge-boson, heavy-quark and Higgs sources through six neutrino flavors and compare the detector-weighted signal with the ten-year IceCube reference~\cite{IceCube}, defining solar compatibility by $\mathcal R_\odot^{\max}<1$ under the stated response variations.

A conditional mixing bound follows after capture is fixed by the recoil signal. For a single cooled population with constant effective volume, define $F_{\rm crit}=2\Gamma_{90}/C_{\rm tot}$, where $\Gamma_{90}$ includes the model neutrino yield. For weakly mixed SHH with $\langle\sigma v\rangle_{11}=a_H F_{H,1}^2$ and $0<F_{\rm crit}<1$,
\begin{equation}
\begin{aligned}
 F_{H,1}&\lesssim
 \sqrt{\frac{V_{\rm eff}}{a_H C_{\rm tot}t_\odot^2}}
 \operatorname{artanh}\sqrt{F_{\rm crit}}\equiv F_{H,1}^{\max},\\
 |\lambda|&\lesssim\frac{\sqrt2\Delta_{HS}}{vB_\beta}
 \sqrt{F_{H,1}^{\max}(1-F_{H,1}^{\max})}.
\end{aligned}
\label{eq:bound}
\end{equation}
Here $B_\beta=|\sin\beta\pm\cos\beta|$ selects the mixed Higgsino combination; BP1 uses the plus sign. Solar limits constrain $\lambda$ at fixed recoil strength, but resonances, additional capture and incomplete cooling modify this bound. Solar compatibility is assessed using the orbital evolution calculation.

\begin{figure*}[!t]
\centering\includegraphics[width=505pt,height=180pt]{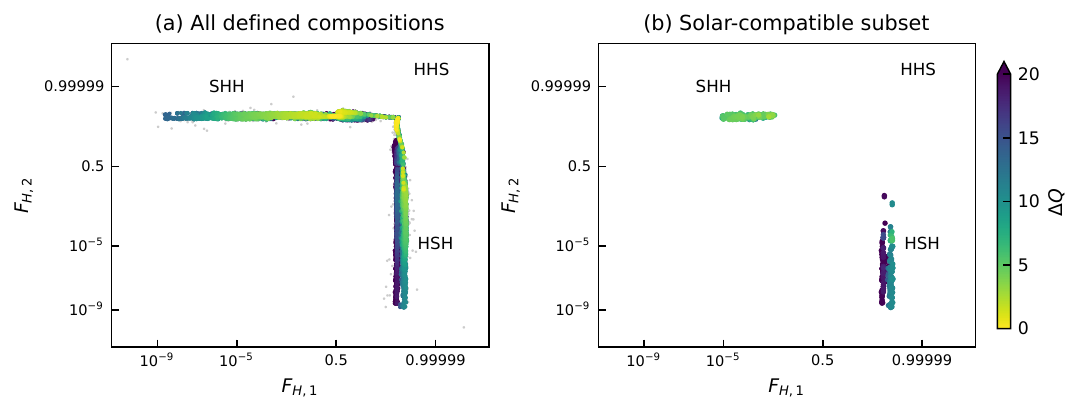}
\caption{Neutralino compositions before (left) and after (right) the ten-year IceCube comparison. All defined compositions appear on the left; nonfinite-$Q$ spectra are gray. All evaluated solar-compatible spectra appear on the right: SHH and HSH survive, while inelastic HHS is excluded under the adopted solar assumptions. Fraction spacing expands nearly pure limits. Colors give total $\Delta Q=Q-Q_{\rm best}$.}
\label{fig:composition}
\end{figure*}

\sectionlead{Composition under the combined constraints}
We scan $m_{\chi_1^0}=200$--$2000\,\GeV$ and $\delta_{21}=100$--$600\,\keV$, varying scalar and coannihilation parameters and both signs of $\mu$. The bino, wino and gluino soft masses are positive and fixed at $3\,\mathrm{TeV}$. Table~\ref{tab:model} lists the requirements defining $G_{\rm model}$ and the additional benchmark collider test. We rank spectra with
\begin{equation}
 Q=Q_{\rm LZ}^{\rm ROI}+2N_{\rm SB}+Q_\gamma+G_{\rm model},
\label{eq:Q}
\end{equation}
where $Q_\gamma$ compares dwarf-galaxy signals~\cite{Fermi}. The accepted recoil spectrum includes coherent inelastic, SI and SD contributions. The normalized score $q_{\rm LZ}=-2\ln[\mathcal L_{\rm LZ}^{\rm prof}/\mathcal L_{\rm LZ,ref}]$ includes the recoil spectrum, calibration profiling and a zero-event sideband with unit acceptance over $350$--$590\,\keV$. This energy-only likelihood approximates the published LZ response.

Figure~\ref{fig:composition} shows that SHH and HSH survive the ten-year IceCube comparison, whereas no evaluated inelastic HHS solution does. BP1 has $F_{H,1}=0.001917$, $\lambda=7.99\times10^{-4}$ and $\mathcal R_\odot=(2.58$--$3.03)\times10^{-3}$. Its high-energy recoil signal coexists with $F_\Gamma\simeq1.1\times10^{-4}$ and $\langle\sigma v\rangle_{11}=1.47\times10^{-31}\,\mathrm{cm}^3\,\mathrm{s}^{-1}$, without imposed cooling suppression. BP4 assumes a thermal core. Within $\Delta Q\le11.83$, solar-compatible SHH spectra reach sampled maxima $F_{H,1}=0.00374$ and $\lambda=0.00111$.

The comparison pair $\mathrm{BP2}^{\rm c}/\mathrm{BP3}^{\rm c}$ has masses, splittings and high-energy yields differing by $0.16\%$, $1.3\%$ and $11.5\%$. The $228$ and $225\,\keV$ gaps match opposite-sign roots, while mixing involves a same-sign Higgsino about $2\,\GeV$ away. This separation permits $\lambda\sim10^{-4}$ despite the residual sub-MeV degeneracy.

Their capture rates, $5.95\times10^{19}$ and $4.97\times10^{19}\,\mathrm{s}^{-1}$, accompany vastly different annihilation efficiencies: $F_\Gamma\sim10^{-12}$ for $\mathrm{BP2}^{\rm c}$ and $0.731$--$0.798$ for $\mathrm{BP3}^{\rm c}$. Weak singlino annihilation and inefficient cooling suppress the former; electroweak Higgsino annihilation sustains the latter. Both sources are dominated by $WW$ and $ZZ$. With $f=2.50\times10^{-5}$, $\mathrm{BP3}^{\rm c}$ survives at $\mathcal R_\odot=0.762$--$0.925$, although its high-energy recoil fit is weaker.

In $\mathrm{BP3}^{\rm c}$, $|C_{13}|^2\simeq0.9992$ lies in the closed transition with a $2.12\,\GeV$ gap. Its low-energy signal contains about $1.27$ elastic events, whereas the $0.214$ events of $\mathrm{BP2}^{\rm c}$ are almost entirely inelastic. Their neutron SI cross sections, $1.72\times10^{-47}$ and $8.69\times10^{-56}\,\mathrm{cm}^2$, expose a complementary ground-state observable, subject to scalar cancellations.

\begin{figure}[!tb]
\centering\includegraphics[width=240pt,height=258pt]{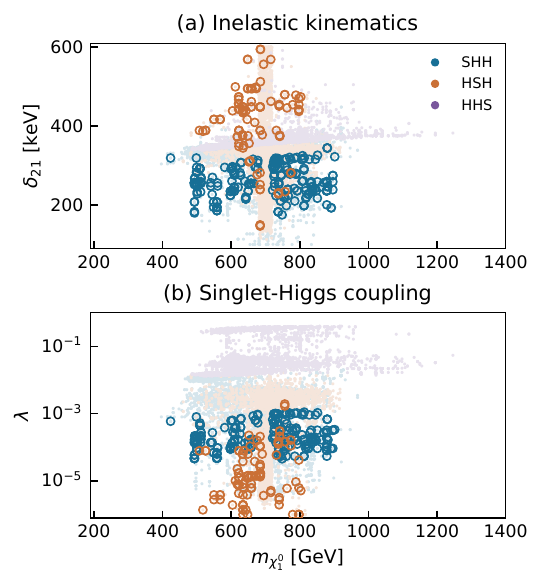}
\caption{Mass dependence of (a) $\delta_{21}$ and (b) the singlet--Higgs coupling $\lambda$ for finite-score spectra with $\Delta Q\le11.83$. Pale filled points distinguish SHH, HSH and HHS; dark open circles mark solar-compatible spectra in both panels.}
\label{fig:parameters}
\end{figure}

Figure~\ref{fig:parameters} shows solar-compatible SHH and HSH masses of $423.5$--$902.0$ and $507.5$--$803.0\,\GeV$ within the selected score window. These sampled intervals reflect coannihilation and the joint recoil--solar comparison. HHS can fit the inelastic recoil well: a tested $528.7\,\GeV$ example has $q_{\rm LZ}=1.515$, yet $\mathcal R_\odot=(3.75$--$4.55)\times10^3$.

These ordering conclusions assume the SHM Maxwellian halo and Helm coherent response. We have not explored loop-induced cancellations that could further suppress elastic cooling. A modified halo tail or sufficiently inhibited cooling can leave captured Higgsinos on extended orbits and weaken the solar constraint~\cite{Higgsino}; such possibilities require a renewed joint recoil--solar analysis.

All four benchmarks satisfy the covered LHC searches. Prompt stau decays remove the charged long-lived signature while preserving the relic abundance (Supplemental Material).

\sectionlead{Elastic scattering with closed transitions}
Although solar constraints exclude inelastic HHS, this ordering may remain viable when elastic scattering dominates. BP4 has a $2.98\,\GeV$ gap, closing coherent upscattering on Earth and in the Sun. Diagonal axial $Z$ exchange survives through $D_1=-0.01014$: a Majorana neutralino has no diagonal vector current, but unequal up/down Higgsino components yield $d_q\propto G_FD_1T_3^q$ and $\sigma_N^{\rm SD}\propto D_1^2$ for heavy squarks. Scalar interference near an SI blind spot~\cite{Blind,Roy} need not force $D_1=0$. BP4 has neutron SD and SI cross sections $3.21\times10^{-42}$ and $7.40\times10^{-49}\,\mathrm{cm}^2$. Signed proton--neutron interference yields $3.28$ low-energy and $0.609$ high-energy xenon events.

\begin{figure}[!tb]
\centering\includegraphics[width=240pt,height=256pt]{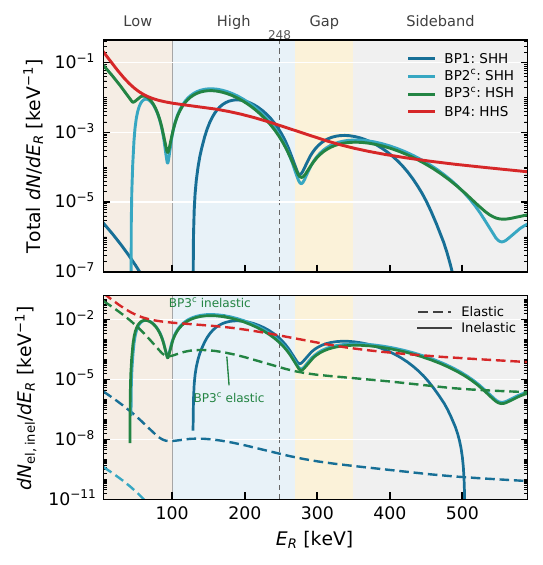}
\caption{Recoil spectra before acceptance at $2.84$ tonne-years: totals (upper), elastic (dashed) and coherent inelastic (solid) components (lower). In the high-energy ROI, $\mathrm{BP3}^{\rm c}$'s green inelastic curve lies above its elastic curve. Shading marks the low-energy ($5.4$--$100\,\keV$) and high-energy ($100$--$269.9\,\keV$) ROI, gap ($269.9$--$350\,\keV$) and sideband ($350$--$590\,\keV$); the vertical line marks $248\,\keV$.}
\label{fig:recoils}
\end{figure}

Figure~\ref{fig:recoils} shows similar high-energy inelastic lobes for the comparison pair. For $\mathrm{BP3}^{\rm c}$, elastic scattering supplies $85.1\%$ of the low-energy yield and coherent inelastic scattering $97.8\%$ of the high-energy yield after acceptance. BP1's narrow signal follows its endothermic threshold and Helm minima; BP4 produces a broad elastic spectrum.

Closing the transition removes efficient heavy-element capture. BP4 annihilates mainly into $WW$ and $ZZ$, with smaller $hZ$ and $t\bar t$ contributions, but elastic capture gives only $4.14\times10^{19}\,\mathrm{s}^{-1}$. Assuming a thermal core, $\Gamma_A\simeq C/2$ and $\mathcal R_\odot=0.508$--$0.777$. Its solar suppression thus comes from reduced capture, rather than the small annihilation efficiency of $\mathrm{BP2}^{\rm c}$.

Elastic scattering may also proceed through a light CP-odd Higgs $a$, whose pseudoscalar couplings favor momentum-dependent nuclear spin scattering~\cite{AxionPortal}. The propagator enhancement saturates for $m_a\ll q$, so reducing its mass alone is insufficient. In the NMSSM, $a$ needs doublet admixture to couple to quarks, while its neutralino coupling depends on $\lambda$, $\kappa$ and composition~\cite{NMSSM}. These parameters also shape the scalar sector. Enhancing the coupling product can overproduce $h_{125}\to aa$ decays or conflict with the Higgs mass, signal strengths, relic density and SI suppression. The key question is whether an $a$-mediated interaction can account for the high-energy event while remaining consistent with the low-energy recoil data. A viable realization may occupy only a very narrow parameter region where cancellations suppress both $h_{125}\to aa$ and SI scattering, while the relic-density, low-energy recoil and light-pseudoscalar search constraints are simultaneously satisfied.

\sectionlead{Experimental tests of composition and scattering}
At fixed $m_\chi$, $\delta_{21}$ and $|C_{12}|^2$, exchanging singlino and Higgsino dominance leaves the leading coherent recoil spectrum unchanged. To test this operator-level degeneracy within physical spectra, we minimize the directional Poisson deviance~\cite{Asimov},
\small
\begin{equation}
 \mathcal D_{A\to B}=\min_{\theta_B\in\mathcal T_B}
 2\sum_k\left[\mu_{B,k}-\mu_{A,k}
 +\mu_{A,k}\ln\frac{\mu_{A,k}}{\mu_{B,k}}\right],
\label{eq:crossfit}
\end{equation}
\normalsize
where the bin means share an exposure and response, and $\mathcal T_B$ contains physical alternatives. Six references admit opposite-ordering high-energy matches with $\mathcal D\sim10^{-5}$--$10^{-4}$. Including the full spectrum and solar selection raises the minimum to $0.93$--$3.27$ after $23\,\keV$ Gaussian broadening (Supplemental Material).

\par\noindent\begin{minipage}{\columnwidth}\makeatletter\def\@captype{table}\makeatother
\caption{Baseline requirements entering $G_{\rm model}$ and the additional SUSY collider test of the benchmarks in Table~\ref{tab:bench}.}
\label{tab:model}
\centering\small\setlength{\tabcolsep}{3pt}\renewcommand{\arraystretch}{1.12}
\begin{tabular}{@{}>{\raggedright\arraybackslash}p{.25\columnwidth}>{\raggedright\arraybackslash}p{.70\columnwidth}@{}}
\toprule
Probe & Physical requirement\\\midrule
Relic abundance & Planck thermal dark-matter density, with $20\%$ theory tolerance~\cite{Planck2018}.\\
\addlinespace[3pt]
Perturbativity & $\lambda^2+\kappa^2\le0.5$ at low scale; one-loop evolution to $2\times10^{16}\,\GeV$ with couplings below $\sqrt{4\pi}$~\cite{NMSSM}.\\
\addlinespace[3pt]
Higgs sector & LEP, Tevatron and LHC exclusions (HiggsBounds); observed Higgs mass and signal strengths (HiggsSignals)~\cite{HiggsBounds5,HiggsSignals2}.\\
\addlinespace[3pt]
Flavor & Rare $B\to X_s\gamma$ and $B_s\to\mu^+\mu^-$ decays~\cite{HFLAV,BsMuon}.\\
\addlinespace[3pt]
Collider searches for SUSY & ATLAS/CMS searches evaluated with the benchmark production rates and decay patterns; $r_{\rm LHC}<1$ in the covered searches~\cite{SModelS,CMSCompressed}.\\
\bottomrule
\end{tabular}
\end{minipage}\par\vspace{12pt}

Low-energy SI scattering probes the ground state. At ten times the exposure, $\mathrm{BP3}^{\rm c}$ predicts $12.1$ low-energy SI events, while $\mathrm{BP2}^{\rm c}$ predicts virtually none; their total low-energy counts are $14.89$ and $2.14$. An excess or its absence can test these representatives. Combining this information with solar data tests HSH more sharply, although competing scalar blind spots must be allowed to refit.

Xenon--tungsten complementarity tests the scattering mechanism~\cite{Frontier,CRESST}. For BP1, $v_\star=\sqrt{2\delta_{21}/\mu_{\chi A}}$ is $765\,\mathrm{km\,s}^{-1}$ on $^{131}$Xe and $671\,\mathrm{km\,s}^{-1}$ on $^{184}$W; at $798\,\mathrm{km\,s}^{-1}$, their lower recoil edges are $143$ and $71\,\keV$. Tungsten accesses a broader endothermic interval. Spin-zero isotopes retain vector scattering; BP4's SD signal involves spinful isotopes, including $^{183}$W ($14.3\%$). A CRESST prediction must include residual SI scattering, CaWO$_4$ composition and acceptance. A coherent tungsten signal with suppressed low-energy xenon recoils would disfavor BP4's axial mechanism; a broad elastic spectrum preserves an HHS possibility despite its inelastic solar exclusion. Mechanism identification must therefore precede ordering exclusion. Identical masses and effective operators remain composition-degenerate across targets. Extended-window xenon measurements, including XLZD~\cite{XLZD}, provide complementary tests.

\sectionlead{Conclusions}
A nuclear recoil can measure a DM transition without determining which state occupies the halo. In our study, the SHH and HSH orderings embody this ambiguity analytically: the same singlino--Higgsino transition can accompany very different diagonal couplings and annihilation rates. Fixing its recoil normalization also fixes the associated solar capture rate for a specified halo, so the remaining solar contrast measures annihilation and orbital evolution. Weak singlino annihilation allows SHH to survive without requiring $\lambda v$ to be smaller than the recoil splitting; HSH can survive when its strong Higgsino transition to the third state is inaccessible. The evaluated inelastic HHS fits fail the ten-year IceCube comparison, while closing their excitation admits a capture-limited elastic alternative. Low-energy SI measurements and xenon--tungsten recoil comparisons consequently address two distinct questions: the identity of the ground state and the mechanism producing the event. Solar neutrinos connect them by probing the annihilation and spatial evolution of the state that actually fills the halo.

\begin{acknowledgments}
\textit{Acknowledgments}---This work is supported by the National Natural Science Foundation of China under Grant No. 12335005, the Natural Science Foundation of Henan Province under Grant No. 252300421771, the Henan Postdoctoral Research Project under Grant No. HN2026054, and the PI Research Fund from Henan Normal University under Grant No. 5101029470335.
\end{acknowledgments}

\onecolumngrid
\clearpage
\makeatletter
\do@output@MVL{\setcounter{page}{1}}
\makeatother
\setcounter{equation}{0}
\setcounter{figure}{0}
\setcounter{table}{0}
\setcounter{section}{0}
\setcounter{footnote}{0}
\renewcommand{\theHsection}{SM.\arabic{section}}
\makeatletter
\let\footnotesize\combined@smfootnotesize
\setcounter{topnumber}{\combined@topnumber}
\setcounter{bottomnumber}{\combined@bottomnumber}
\setcounter{dbltopnumber}{\combined@dbltopnumber}
\let\topfraction\combined@topfraction
\let\bottomfraction\combined@bottomfraction
\let\textfraction\combined@textfraction
\let\dbltopfraction\combined@dbltopfraction
\frontmatter@init
\global\setbox\absbox\box\voidb@x
\let\author\combined@author
\let\affiliation\combined@affiliation
\let\email\combined@email
\let\thanks\combined@thanks
\let\and\combined@and
\let\maketitle\frontmatter@maketitle
\patchcmd{\titleblock@produce}{\label{FirstPage}}{\label{SM:FirstPage}}{}{\PackageError{combined}{Cannot namespace the supplemental title label}{Use REVTeX 4.2.}}
\patchcmd{\endthebibliography}{\label{LastBibItem}}{\label{SM:LastBibItem}}{}{\PackageError{combined}{Cannot namespace the supplemental bibliography label}{Use REVTeX 4.2.}}
\patchcmd{\frontmatter@footnotemark}{frontmatter.}{SM.frontmatter.}{}{\PackageError{combined}{Cannot namespace supplemental author links}{Use REVTeX 4.2.}}
\patchcmd{\present@bibnote}{frontmatter.}{SM.frontmatter.}{}{\PackageError{combined}{Cannot namespace supplemental author notes}{Use REVTeX 4.2.}}
\makeatother

\title{\texorpdfstring{Supplemental Material\\Can Nuclear Recoils and Solar Neutrinos Resolve the Composition of Singlino-Higgsino Dark Matter?}{Supplemental Material: Can Nuclear Recoils and Solar Neutrinos Resolve the Composition of Singlino-Higgsino Dark Matter?}}
\author{Jingwei Lian}
\email{lianjw@hist.edu.cn}
\affiliation{Henan Institute of Science and Technology, Xinxiang 453003, P. R. China}
\affiliation{Centre for Theoretical Physics, Henan Normal University, Xinxiang 453007, P. R. China}
\author{Jin Min Yang}
\email{jmyang@itp.ac.cn}
\affiliation{Centre for Theoretical Physics, Henan Normal University, Xinxiang 453007, P. R. China}
\affiliation{Institute of Theoretical Physics, Chinese Academy of Sciences, Beijing 100190, P. R. China}
\maketitle
\setcounter{equation}{0}
\renewcommand{\theequation}{S\arabic{equation}}
\renewcommand{\theHequation}{S\arabic{equation}}
\renewcommand{\thetable}{S\arabic{table}}
\renewcommand{\theHtable}{S\arabic{table}}
\renewcommand{\thefigure}{S\arabic{figure}}
\renewcommand{\theHfigure}{S\arabic{figure}}
\renewcommand{\thesection}{S\arabic{section}}

\section{Neutralino mixing and \texorpdfstring{$Z$}{Z} couplings}

In the three-state singlino--Higgsino approximation, the CP-conserving neutralino mass matrix in the basis $(\widetilde H_d^0,\widetilde H_u^0,\widetilde S)$ is~\cite{SM:NMSSM}
\begin{equation}
 M_3=
 \begin{pmatrix}
 0&-\mu&-\lambda v\sin\beta\\
 -\mu&0&-\lambda v\cos\beta\\
 -\lambda v\sin\beta&-\lambda v\cos\beta&m_S
 \end{pmatrix},
 \label{SM:eq:massmatrix}
\end{equation}
where $m_S=2\kappa\mu/\lambda$. Its signed eigenvalues $t_i$ correspond to physical masses $m_i=|t_i|$~\cite{SM:Spinors} and satisfy the characteristic equation $\det(tI-M_3)=0$,
\begin{equation}
 (t-m_S)(t^2-\mu^2)
 -\lambda^2v^2(t-\mu\sin2\beta)=0.
 \label{SM:eq:cubic}
\end{equation}

For $N_{iS}\ne0$ and $t_i^2\ne\mu^2$, solving the first two rows of the eigenvalue equation gives the Higgsino-to-singlino component ratios,
\begin{equation}
 \begin{aligned}
 \frac{N_{id}}{N_{iS}}
 &=
 \frac{\lambda v(\mu\cos\beta-t_i\sin\beta)}
      {t_i^2-\mu^2},\\
 \frac{N_{iu}}{N_{iS}}
 &=
 \frac{\lambda v(\mu\sin\beta-t_i\cos\beta)}
      {t_i^2-\mu^2}.
 \end{aligned}
\end{equation}
With $F_{H,i}=|N_{id}|^2+|N_{iu}|^2$ and $|N_{iS}|^2=1-F_{H,i}$, normalization then gives~\cite{SM:Blind}
\begin{equation}
 \begin{aligned}
 \frac{F_{H,i}}{1-F_{H,i}}
 &=
 \left|\frac{N_{id}}{N_{iS}}\right|^2
 +\left|\frac{N_{iu}}{N_{iS}}\right|^2\\
 &=
 \frac{\lambda^2v^2
 (t_i^2+\mu^2-2t_i\mu\sin2\beta)}
 {(\mu^2-t_i^2)^2}.
 \end{aligned}
 \label{SM:eq:FH}
\end{equation}

The component ratios also give explicit expressions for the neutralino $Z$ couplings. Writing $P_i=t_i^2+\mu^2-2\mu t_i\sin2\beta$ and $D_i=C_{ii}$, we obtain
\begin{equation}
 \begin{aligned}
 \frac{D_i}{F_{H,i}}
 &=
 \frac{(\mu^2-t_i^2)\cos2\beta}{P_i},\\
 \frac{|C_{ij}|^2}{F_{H,i}F_{H,j}}
 &=
 \frac{(\mu^2-t_it_j)^2\cos^22\beta}{P_iP_j}.
 \end{aligned}
 \label{SM:eq:Zcouplings}
\end{equation}
These relations separate the dependence on the total Higgsino fractions from the relative weights of the two Higgsino components. For fixed $\tan\beta\ne1$, the limit $t_i\to\mu$ and $t_j\to-\mu$ gives $D_i/F_{H,i}\to0$ and $D_j/F_{H,j}\to0$, whereas $|C_{ij}|^2/(F_{H,i}F_{H,j})\to1$. Thus, balancing the two Higgsino components suppresses the diagonal axial couplings while allowing the transition strength to approach its maximal value for the given Higgsino fractions.

\section{SHH, HSH and HHS splitting relations}

For $\mu>0$, the combinations $H_A=(\widetilde H_d^0-\widetilde H_u^0)/\sqrt2$ and $H_B=(\widetilde H_d^0+\widetilde H_u^0)/\sqrt2$ diagonalize the Higgsino part of Eq.~(\ref{SM:eq:massmatrix}), with signed eigenvalues $+\mu$ and $-\mu$. Taking the difference and sum of its singlino mixing entries gives
\begin{equation}
 \begin{aligned}
 a&=\frac{\lambda v(\cos\beta-\sin\beta)}{\sqrt2},\\
 b&=-\frac{\lambda v(\sin\beta+\cos\beta)}{\sqrt2}.
 \end{aligned}
 \label{SM:eq:mixingentries}
\end{equation}
The full three-state matrix in this basis is therefore
\begin{equation}
 M_{ABS}=
 \begin{pmatrix}
 \mu&0&a\\
 0&-\mu&b\\
 a&b&m_S
 \end{pmatrix}.
 \label{SM:eq:rotatedmatrix}
\end{equation}
For $m_S\simeq-\mu$, the singlino is close in signed mass to $H_B$, while its separation from $H_A$ is approximately $2\mu$. When $|a|\ll|\mu-m_S|$, the small $H_A$ admixture permits a two-state treatment of $H_B$--singlino mixing. This treatment allows shifts of the Higgsino diagonal masses, including bino and wino corrections, while neglecting mixing with the remaining Higgsino.

Writing $m_i=m_{\tilde{\chi}_i^0}$ for the ordered physical masses, the two relevant mass differences are
\begin{equation}
 \begin{aligned}
 \delta_{21}&=m_2-m_1,\\
 \Delta_{HS}&=
 \begin{cases}
 m_3-m_1,&\mathrm{SHH},\\
 m_3-m_2,&\mathrm{HSH}.
 \end{cases}
 \end{aligned}
 \label{SM:eq:massgaps}
\end{equation}
Within the two-state approximation, $f=|C_{12}|^2$ equals the Higgsino fraction of the singlino-like state: $F_{H,1}$ in SHH and $F_{H,2}$ in HSH. Diagonalization gives
\begin{equation}
 b^2=f(1-f)\Delta_{HS}^2.
 \label{SM:eq:twostatemixing}
\end{equation}
The singlino diagonal entry is the mixing-weighted sum of the two negative eigenvalues,
\begin{equation}
 m_S\simeq
 \begin{cases}
 -m_1-f\Delta_{HS},&\mathrm{SHH},\\
 -m_2-f\Delta_{HS},&\mathrm{HSH}.
 \end{cases}
 \label{SM:eq:singlinomass}
\end{equation}
Thus $f\Delta_{HS}$ is the mixing-induced shift that must be included when choosing $m_S$ to obtain the required $\delta_{21}$. The remaining Higgsino lies between the two mixed states in SHH and below both in HSH. Its mass corrections therefore matter for the ordering and the small recoil splitting, even when their effect on the mixing fractions is small.

For $\mu<0$ and $m_S\simeq\mu$, the singlino instead mixes mainly with $H_A$. Equation~(\ref{SM:eq:twostatemixing}) then holds with $a$ replacing $b$, yielding
\begin{equation}
 |\lambda|\simeq
 \frac{\sqrt2\Delta_{HS}\sqrt{f(1-f)}}
 {v|\cos\beta-\sin\beta|}.
 \label{SM:eq:negative_mu_mixing}
\end{equation}
This relation applies to the negative-$\mu$ HSH comparison $\mathrm{BP3}^{\rm c}$; BP1 has $\mu>0$ and uses the $H_B$ mixing entry $b$ in Eq.~(\ref{SM:eq:mixingentries}). At $\tan\beta=1$, $a$ vanishes, so this approximation cannot sustain a nonzero $f$ at fixed $\Delta_{HS}$.

In HHS, the singlino-like state is heavier than both Higgsinos. Away from $m_S=\pm\mu$, singlino mixing induces second-order shifts in the signed Higgsino eigenvalues,
\begin{equation}
 \begin{aligned}
 \Delta t_+&\simeq\frac{a^2}{\mu-m_S}
 =\frac{\lambda^2v^2(1-\sin2\beta)}{2(\mu-m_S)},\\
 \Delta t_-&\simeq\frac{b^2}{-\mu-m_S}
 =-\frac{\lambda^2v^2(1+\sin2\beta)}{2(\mu+m_S)}.
 \end{aligned}
 \label{SM:eq:HHSshifts}
\end{equation}
These are tree-level corrections quadratic in the mixing entries. Including the leading bino and wino contribution gives
\begin{equation}
 \delta_{21}\simeq
 \left|G_0+\frac{\lambda^2v^2(m_S-\mu\sin2\beta)}
 {m_S^2-\mu^2}\right|,
 \label{SM:eq:HHSsplitting}
\end{equation}
where $G_0=m_Z^2(s_W^2/M_1+c_W^2/M_2)$. The analytic expressions above clarify the origin of the mass splittings. Our numerical analysis uses the full five-state neutralino spectrum, including the one-loop mass corrections implemented in SARAH/SPheno~\cite{SM:SARAH,SM:SPheno}.
The singlino mass parameter is adjusted using these loop-corrected masses to realize the sub-MeV benchmark splittings.
\section{Common recoil response and energy windows}
The isotope-summed coherent rate is
\begin{equation}
 \frac{dR_Z}{dE_R}=\sum_A\frac{\xi_A\rho_\chi}{2m_\chi\mu_{\chi n}^2}
 \sigma_n^Z[(A-Z)+Zr_Z]^2F_A^2(q)\,\eta(v_{\min}),
\end{equation}
with mass fractions $\xi_A$, $r_Z=-(1-4s_W^2)$ and the conventional normalization $\sigma_n^Z=G_F^2\mu_{\chi n}^2 f/(2\pi)$. This normalization specifies the transition amplitude; it is not a zero-speed endothermic cross section. The mean inverse speed $\eta$ uses the truncated Maxwellian specified in the Letter.

All coherent $Z$ and elastic SI rates use the Helm form factor,
\begin{equation}
\begin{aligned} F_A(q)&=\frac{3j_1(qR_1)}{qR_1}\exp[-(qs)^2/2],\\
 R_1^2&=c^2+\frac{7\pi^2a^2}{3}-5s^2,\end{aligned}
\end{equation}
where $s=0.9\,\mathrm{fm}$, $a=0.52\,\mathrm{fm}$ and $c=(1.23A^{1/3}-0.60)\,\mathrm{fm}$. The scalar rate retains the actual signed ratio $f_p/f_n$, which differs from $r_Z$.

The SD response is evaluated with the GCN5082 one-body matrix~\cite{SM:PandaXSD,SM:Anand} for the two spin-active xenon isotopes, $^{129}$Xe and $^{131}$Xe. Its transverse and longitudinal spin responses are retained in the signed proton--neutron basis:
\begin{equation}
 \sigma_A^{\rm SD}(q)=\frac4\pi\mu_{\chi A}^2
 \begin{pmatrix}a_p&a_n\end{pmatrix}
 \mathcal F_A^{\rm spin}(q)
 \begin{pmatrix}a_p\\a_n\end{pmatrix}.
\end{equation}
The free-proton normalization is $\mathcal F_p^{\rm spin}(0)=3$, giving $\sigma_p^{\rm SD}=12\mu_{\chi p}^2a_p^2/\pi$. The nuclear response satisfies positivity and the zero-momentum transverse/longitudinal normalization. A neutron-only exclusion curve cannot substitute for this correlated nuclear prediction.

The exposure is $2.84$ tonne-years. The full energy-only recoil comparison integrates the accepted true-energy rate over $0.1$--$300\,\keV$, folding the published LZ acceptance~\cite{SM:LZ}. The nominal energy interval in Fig.~\ref{fig:recoils} of the Letter ends at $269.9\,\keV$, where the published efficiency crosses $50\%$; this is not a hard true-energy cutoff. Table~\ref{tab:bench} projects accepted counts into $5.4$--$100$ and $100$--$269.9\,\keV$. Contributions below $5.4\,\keV$ and the finite efficiency tail above $269.9\,\keV$ remain in the full recoil normalization. The $269.9$--$350\,\keV$ gap is displayed separately; the sideband is the raw $350$--$590\,\keV$ rate with unit acceptance. Its event expectation is multiplied by two in the zero-count contribution to $-2\ln\mathcal L$. These energy projections approximate the detector selection and do not reproduce its full S1/S2 response.

For a normalized accepted true-energy density $p(E)$, statistical resolution $\sigma_{\rm stat}=23\,\keV$ and calibration displacement $b_E$, the conditional density at the candidate is
\begin{equation}
 p_{248}(b_E)=\int dE\,p(E)\,
 \frac{e^{-(248-E-b_E)^2/(2\sigma_{\rm stat}^2)}}
 {\sqrt{2\pi}\sigma_{\rm stat}}.
\end{equation}
With signal expectation $N$, residual background density $\beta$ and $\sigma_{\rm cal}=23\,\keV$, we minimize
\begin{equation}
 Q_{\rm LZ}^{\rm ROI}=2N-2\ln[\beta+Np_{248}(b_E)]
 +(b_E/\sigma_{\rm cal})^2,
\end{equation}
using $\beta=6.3789911\times10^{-6}\,\keV^{-1}$. Including the zero-count sideband factor, the profiled recoil likelihood is proportional to
\begin{equation}
\begin{aligned}
 \mathcal L_{\rm LZ}^{\rm prof}(\theta)\propto
 \max_{b_E}\bigl\{&[\beta+Np_{248}(b_E)]\\
 &\times e^{-N-N_{\rm SB}-b_E^2/(2\sigma_{\rm cal}^2)}\bigr\}.
\end{aligned}
\end{equation}
We define the reference-normalized comparison by
\begin{equation}
\begin{aligned}
 q_{\rm LZ}&=-2\ln\frac{\mathcal L_{\rm LZ}^{\rm prof}(\theta)}
 {\mathcal L_{\rm LZ,ref}}\\
 &=Q_{\rm LZ}^{\rm ROI}+2N_{\rm SB}-Q_{\rm LZ,ref}.
\end{aligned}
\end{equation}
The numerator and reference use identical density and nuisance conventions. We adopt $Q_{\rm LZ,ref}=11.17161584$, evaluated with the Helm response and the same sideband term. This reference fixes the additive normalization of $q_{\rm LZ}$ and is not assumed to be the global NMSSM minimum. Its value does not affect likelihood differences or spectrum rankings. 

\section{Parameter domain and representative spectra}
The parameter ranges in Table~\ref{SM:tab:domain} define the sampling domain rather than physical exclusions. At a renormalization scale of $2\,\mathrm{TeV}$, the bino, wino and gluino soft masses are fixed at $+3\,\mathrm{TeV}$, and first- and second-generation squark soft masses at $5\,\mathrm{TeV}$. The elastic comparison releases the targeted sub-MeV splitting. Spectra, relic densities and nucleon amplitudes are evaluated with SARAH/SPheno and micrOMEGAs~\cite{SM:SARAH,SM:SPheno,SM:MO}. The particle-physics requirements are summarized in Table~\ref{tab:model} of the Letter. The four reported benchmarks additionally satisfy the covered collider searches; this qualification is not inferred for every projected scan point.

Figure~\ref{fig:composition} of the Letter shows all defined neutralino compositions in the left panel and the evaluated solar-compatible subset in the right panel. Both panels use a common $Q_{\rm best}=10.21130179$, with the color scale saturating at $\Delta Q=20$. The composition axes expand the nearly pure limits while labeling the physical fractions. Figure~\ref{fig:parameters} shows finite-score spectra with $\Delta Q\le11.83$ and highlights the solar-compatible SHH and HSH solutions. The surviving mass intervals characterize this sampled score window rather than ordering-specific solar exclusion boundaries.
For the representative benchmarks, Table~\ref{SM:tab:inputs} lists the weak-scale inputs, physical masses and dark-matter observables. Slepton soft inputs are distinguished from the lightest charged-slepton pole mass. Table~\ref{SM:tab:components} gives the recoil contributions in the three experimental energy windows.

\begin{table*}[t]
\caption{Weak-scale inputs, physical spectrum and dark-matter observables of Table~\ref{tab:bench}. Dimensionful inputs and masses are in GeV; nucleon cross sections are in $\mathrm{cm}^2$. $m_S$ and $\mu$ are signed inputs, while all physical masses are positive. $m_{\tilde\ell_1}$ denotes the lightest charged slepton. Additional digits in the neutralino masses specify the sub-MeV splitting.}
\label{SM:tab:inputs}\centering\small\setlength{\tabcolsep}{5pt}\renewcommand{\arraystretch}{1.02}
\begin{tabular}{@{}lrrrr@{}}\toprule
 & BP1 & $\mathrm{BP2}^{\rm c}$ & $\mathrm{BP3}^{\rm c}$ & BP4\\\midrule
\multicolumn{5}{@{}l}{\textit{Weak-scale inputs}}\\[2pt]
$\mu$ & $4.8821\times10^{2}$ & $7.2808\times10^{2}$ & $-7.0963\times10^{2}$ & $-4.8778\times10^{2}$\\
$m_S$ & $-4.9862\times10^{2}$ & $-7.3891\times10^{2}$ & $-7.3771\times10^{2}$ & $-1.1017\times10^{3}$\\
$\lambda$ & $7.9940\times10^{-4}$ & $9.2319\times10^{-5}$ & $1.3389\times10^{-4}$ & $4.4175\times10^{-1}$\\
$\kappa$ & $-4.0823\times10^{-4}$ & $-4.6846\times10^{-5}$ & $6.9592\times10^{-5}$ & $4.9884\times10^{-1}$\\
$\tan\beta$ & $7.3803\times10^{0}$ & $5.5410\times10^{0}$ & $3.1711\times10^{0}$ & $4.4955\times10^{0}$\\
$m_A$ & $2.4960\times10^{3}$ & $2.8274\times10^{3}$ & $4.4987\times10^{3}$ & $2.4206\times10^{3}$\\
$A_\kappa$ & $7.7442\times10^{2}$ & $5.0163\times10^{2}$ & $3.6500\times10^{2}$ & $6.0993\times10^{2}$\\
$A_\lambda$ & $1.9472\times10^{3}$ & $2.2885\times10^{3}$ & $-7.8112\times10^{3}$ & $-1.9952\times10^{3}$\\
$m_{\tilde\ell,\mathrm{soft}}$ & $4.9057\times10^{2}$ & $8.9875\times10^{2}$ & $7.6067\times10^{2}$ & $4.9383\times10^{2}$\\
$m_{\tilde\tau_R,\mathrm{soft}}$ & $4.9757\times10^{2}$ & $8.9875\times10^{2}$ & $7.6067\times10^{2}$ & $4.9659\times10^{2}$\\
$M_{\tilde t}$ & $2.8445\times10^{3}$ & $2.3098\times10^{3}$ & $4.8705\times10^{3}$ & $3.9803\times10^{3}$\\
$X_t/M_{\tilde t}$ & $1.7548\times10^{0}$ & $2.6820\times10^{0}$ & $2.4074\times10^{0}$ & $1.5255\times10^{0}$\\
\addlinespace[5pt]\midrule\addlinespace[3pt]\multicolumn{5}{@{}l}{\textit{Physical masses [GeV]}}\\[2pt]
$m_{\chi_1^0}$ & $498.618084313$ & $738.907544552$ & $737.714342774$ & $504.277813438$\\
$m_{\chi_2^0}$ & $498.618403584$ & $738.907772514$ & $737.714567774$ & $507.261367228$\\
$m_{\chi_3^0}$ & $501.123117742$ & $741.234049762$ & $739.831202627$ & $1097.424965532$\\
$m_{\chi_1^\pm}$ & $4.9998\times10^{2}$ & $7.4010\times10^{2}$ & $7.3944\times10^{2}$ & $5.0645\times10^{2}$\\
$m_{\tilde\ell_1}$ & $5.0343\times10^{2}$ & $9.0588\times10^{2}$ & $7.6266\times10^{2}$ & $5.0649\times10^{2}$\\
$m_{h_1}$ & $1.2394\times10^{2}$ & $1.2239\times10^{2}$ & $1.2227\times10^{2}$ & $1.2345\times10^{2}$\\
$m_{h_2}$ & $2.3570\times10^{2}$ & $6.0055\times10^{2}$ & $6.3999\times10^{2}$ & $9.1356\times10^{2}$\\
$m_{h_3}$ & $2.5469\times10^{3}$ & $2.8587\times10^{3}$ & $4.4058\times10^{3}$ & $2.3822\times10^{3}$\\
$m_{a_1}$ & $7.6106\times10^{2}$ & $7.4564\times10^{2}$ & $6.3553\times10^{2}$ & $9.8641\times10^{2}$\\
$m_{a_2}$ & $2.5468\times10^{3}$ & $2.8586\times10^{3}$ & $4.4054\times10^{3}$ & $2.3803\times10^{3}$\\
$m_{H^\pm}$ & $2.5463\times10^{3}$ & $2.8591\times10^{3}$ & $4.4150\times10^{3}$ & $2.3781\times10^{3}$\\
\addlinespace[5pt]\midrule\addlinespace[3pt]\multicolumn{5}{@{}l}{\textit{Dark-matter observables}}\\[2pt]
$\Omega_\chi h^2$ & $1.2068\times10^{-1}$ & $1.0015\times10^{-1}$ & $1.3127\times10^{-1}$ & $1.1972\times10^{-1}$\\
$\sigma_p^{\mathrm{SI}}$ & $3.2019\times10^{-52}$ & $8.4860\times10^{-56}$ & $1.6799\times10^{-47}$ & $7.7094\times10^{-49}$\\
$\sigma_n^{\mathrm{SI}}$ & $3.2820\times10^{-52}$ & $8.6910\times10^{-56}$ & $1.7183\times10^{-47}$ & $7.3965\times10^{-49}$\\
$\sigma_p^{\mathrm{SD}}$ & $4.6916\times10^{-48}$ & $3.6426\times10^{-52}$ & $1.1381\times10^{-43}$ & $4.1910\times10^{-42}$\\
$\sigma_n^{\mathrm{SD}}$ & $3.5874\times10^{-48}$ & $2.7853\times10^{-52}$ & $8.7027\times10^{-44}$ & $3.2047\times10^{-42}$\\
\bottomrule\end{tabular}\end{table*}

At freeze-out, Higgsino and singlino states coannihilate in both SHH and HSH. In chemical equilibrium, neglecting singlino annihilation and other coannihilators, $\langle\sigma v\rangle_{\rm eff}\simeq[G_H/(G_H+G_S)]^2\langle\sigma v\rangle_H$, with $G_H\simeq8$ and $G_S\simeq2$. The singlino thus dilutes the pure thermal-Higgsino mass scale near $1.1\,\mathrm{TeV}$. Sleptons add Boltzmann-weighted states and annihilation channels and must be included dynamically. These effects explain conditional mass scales rather than universal ordering endpoints; conversion-driven freeze-out would require relaxing the assumed chemical equilibrium.

\begin{table*}[t]
\caption{Accepted low/high-energy and raw sideband counts decomposed into SI, SD and coherent vector contributions, for an exposure of $2.84$ tonne-years and the energy windows of Table~\ref{tab:bench}.}
\label{SM:tab:components}\centering\small\setlength{\tabcolsep}{6pt}
\begin{tabular}{llrrr}\toprule
Point & Window & $N_{\rm SI}$ & $N_{\rm SD}$ & $N_Z$\\\midrule
BP1 & Low & $3.3710\times10^{-5}$ & $3.6277\times10^{-6}$ & $0$\\
BP1 & High & $3.5539\times10^{-7}$ & $6.8537\times10^{-7}$ & $5.2772\times10^{-1}$\\
BP1 & Sideband & $5.1764\times10^{-9}$ & $4.2757\times10^{-8}$ & $3.0447\times10^{-2}$\\
$\mathrm{BP2}^{\rm c}$ & Low & $6.0921\times10^{-9}$ & $1.9297\times10^{-10}$ & $2.1399\times10^{-1}$\\
$\mathrm{BP2}^{\rm c}$ & High & $7.0648\times10^{-11}$ & $4.0156\times10^{-11}$ & $1.3730\times10^{0}$\\
$\mathrm{BP2}^{\rm c}$ & Sideband & $1.2738\times10^{-12}$ & $3.2339\times10^{-12}$ & $4.3841\times10^{-2}$\\
$\mathrm{BP3}^{\rm c}$ & Low & $1.2070\times10^{0}$ & $6.0389\times10^{-2}$ & $2.2140\times10^{-1}$\\
$\mathrm{BP3}^{\rm c}$ & High & $1.3993\times10^{-2}$ & $1.2563\times10^{-2}$ & $1.1889\times10^{0}$\\
$\mathrm{BP3}^{\rm c}$ & Sideband & $2.5211\times10^{-4}$ & $1.0109\times10^{-3}$ & $3.7144\times10^{-2}$\\
BP4 & Low & $7.7224\times10^{-2}$ & $3.2060\times10^{0}$ & $0$\\
BP4 & High & $8.1690\times10^{-4}$ & $6.0780\times10^{-1}$ & $0$\\
BP4 & Sideband & $1.1991\times10^{-5}$ & $3.8269\times10^{-2}$ & $0$\\
\bottomrule\end{tabular}\end{table*}

\begin{table}[t]
\caption{Ranges of the sampled inelastic domain. They specify search coverage, not imposed physical limits. $g_\ell$ is the targeted slepton--LSP mass gap divided by the target LSP mass; the spectrum calculation determines the physical gap. $M_{\tilde t}$ is the common stop soft mass and $X_t=A_t-\mu\cot\beta$.}
\label{SM:tab:domain}\centering\small\setlength{\tabcolsep}{5pt}\renewcommand{\arraystretch}{1.08}
\begin{tabular}{@{}lr@{}}\toprule
Parameter & Sampled interval\\\midrule
$m_\chi$ [GeV] & $200$--$2000$\\
$\delta_{21}$ [keV] & $100$--$600$\\
$\lambda$ & $10^{-6}$--$1/\sqrt2$\\
$\tan\beta$ & $1$--$60$\\
$m_A$ [GeV] & $200$--$6000$\\
$A_\kappa$ [GeV] & $-5000$--$5000$\\
$g_\ell$ & $0.001$--$1$\\
$M_{\tilde t}$ [TeV] & $1$--$5$\\
$X_t/M_{\tilde t}$ & $-3$--$3$\\
\bottomrule\end{tabular}\end{table}

The likelihood uses the full accepted $0.1$--$300\,\keV$ rate, including contributions below $5.4\,\keV$ and above $269.9\,\keV$ that are absent from the low/high-band sums in Table~\ref{tab:bench}. BP4 represents the elastic-scattering scenario and yields $q_{\rm LZ}=9.87139$.

At low energy, BP1 has a weak elastic tail set by its small scalar amplitudes and $|D_1|=1.07\times10^{-5}$. Its coherent endothermic channel opens only in the high-speed halo tail: $v_\star\simeq765\,\mathrm{km\,s}^{-1}$ on $^{131}$Xe, compared with the adopted maximum laboratory speed of $798\,\mathrm{km\,s}^{-1}$. Above threshold the spectrum is proportional to $F_A^2(q)\eta(v_{\min})$. The first two Helm zeros for $^{131}$Xe lie near $95$ and $280\,\keV$; the latter separates the high-energy lobes of BP1. Summing xenon isotopes broadens the diffraction minima, while the SD contribution fills the low-energy elastic minimum. The weak tail and the lobes thus reflect different interactions within the same spectrum.

\section{Transition-normalized solar evolution}
Let $\mathcal F_i(z,t)$ describe the captured particles of state $i$ over orbital phase space. Their evolution has the schematic form
\begin{equation}
 \partial_t\mathcal F_i=S_i+
 \sum_j(\mathcal W_{ij}\mathcal F_j-\mathcal W_{ji}\mathcal F_i)
 +\mathcal D_i-\mathcal E_i-\mathcal A_i.
\end{equation}
The terms describe capture, excitation/de-excitation and elastic scattering, decay, escape, and annihilation. Moving solar targets are included. The annihilation rate is
\begin{equation}
 \Gamma_A=\frac12\sum_{ij}\int d^3r\,n_i(r)n_j(r)
 \langle\sigma v\rangle_{ij},
\end{equation}
where the sum includes both $ij$ and $ji$. Inelastic cooling can stop before a thermal core is reached, so a central Maxwellian population cannot be assumed for every point. The $11$, $12$ and $22$ annihilation channels are retained, with low-speed rates evaluated at $v_{\rm rel}=10^{-4}c$. Three-body excited-state decay is included. Complete velocity averaging, a model-dependent radiative width and Sommerfeld corrections where relevant can modify this approximation.

The orbital collision operator integrates over target velocities, scattering kinematics, solar radius and isotope abundances. The resulting phase-space evolution determines the radial overlap of the captured populations and hence their annihilation rate. Neutrino propagation includes six flavors and the spatial and angular distribution of the source.

Numerical stability is assessed with four orbital evaluations for each of BP1 and $\mathrm{BP3}^{\rm c}$. For BP1, varying the energy grid (316 or 599 nodes), angular-momentum grid (five or seven nodes) and sampling seed, with 1024 scattering samples, changes the annihilation rate and solar ratio by about $6\%$. Median annihilation and number radii are $0.0085R_\odot$ and $0.034R_\odot$, respectively, illustrating why particle number alone does not fix the neutrino source. BP4 instead assumes a thermal core, with moving-hydrogen capture and doubled neutrino-energy resolution as checks.

The same transition strength $f$ multiplies $K_{\rm Xe}$ and $K_\odot$ for the specified coherent operator. Eliminating it yields $C_Z=N_ZK_\odot/K_{\rm Xe}$. This relation is conditional on masses, isotope/nuclear response and halo shape. Other coherent transitions and elastic channels add capture that is not determined by $N_Z$. Scaling the local density multiplies both kernels, while changing the halo shape changes their ratio because terrestrial recoils and solar capture select different incident speeds.

For a single cooled population with constant capture and negligible evaporation,
\begin{equation}
 \dot N=C-C_AN^2,\qquad
 \Gamma_A=\frac C2\tanh^2(t_\odot\sqrt{CC_A}).
\end{equation}
A detector-weighted limit $\Gamma_{90}$ gives, for $2\Gamma_{90}<C$,
\begin{equation}
 C_A\le\frac1{Ct_\odot^2}
 \left[\operatorname{artanh}\sqrt{\frac{2\Gamma_{90}}C}\right]^2.
\end{equation}
If $C\le2\Gamma_{90}$, even the capture-limited equilibrium source is below the limit. This is the relevant limiting mechanism for elastic BP4, rather than a small Higgsino self-annihilation rate.

For $C=C_0f$, $C_A=a_Hf^2/V_{\rm eff}$ and $t_\odot\sqrt{CC_A}\ll1$, the weakly mixed SHH source is
\begin{equation}
 \Gamma_A^{\rm SHH}\simeq
 \frac{C_0^2a_Ht_\odot^2}{2V_{\rm eff}}f^4.
\end{equation}
HSH instead retains $C_A\simeq a_H/V_{\rm eff}$ and has $\Gamma_A\propto f^2$ in the same dilute limit. At fixed $\Delta_{HS}$ and $\beta$, weak mixing gives $f\propto\lambda^2$ and hence $\Gamma_A^{\rm SHH}\propto\lambda^8$. This trajectory reduces both the recoil signal and capture. At a fixed measured transition signal, capture is fixed and the SHH source scales with its annihilation coefficient, approximately as $f^2$ or $\lambda^4$ if the same spatial distribution is maintained.

For weakly mixed SHH at fixed capture and a specified effective volume, the cooled-population inequality gives the Letter's $F_{H,1}^{\max}$ when $C_A=a_H F_{H,1}^2/V_{\rm eff}$. The two-state mixing relation then yields
\begin{equation}
 |\lambda|\simeq\frac{\sqrt2\Delta_{HS}}{v|\sin\beta\pm\cos\beta|}
 \sqrt{F_{H,1}(1-F_{H,1})}.
\end{equation}
For $F_{H,1}<1/2$ this relation is monotonic, so substituting $F_{H,1}^{\max}$ gives the conditional upper bound. The plus sign applies to the positive-$\mu$ SHH configuration of BP1; the minus sign selects the other Higgsino combination. A varying source volume, resonant annihilation or additional capture invalidates a universal numerical upper limit. In equilibrium, both orderings approach $C/2$, and this particular annihilation contrast loses sensitivity.

For the negative-$\mu$ HSH comparison, $|\lambda|\simeq\sqrt2\Delta_{HS}\sqrt{f}/[v|\sin\beta-\cos\beta|]$. Approaching $\tan\beta=1$ can reduce the SM-like Higgs amplitude but also reduces this mixing entry. Maintaining a fixed accessible transition then requires increasing $|\lambda|$ or changing $\Delta_{HS}$. The simplest Higgs blind limit therefore does not automatically erase the solar--elastic distinction. Additional scalar interference and signed gaugino terms must be tested in the full model.

The detector-weighted comparison to IceCube~\cite{SM:IceCube} is
\begin{equation}
 \mathcal R_\odot=\frac{\Gamma_A}{\Gamma_{90}^{WW}(m_\chi)}
 \frac{Y_{\rm det}^{\rm model}}{Y_{\rm det}^{WW,\rm ref}}.
\end{equation}
Six-flavor neutrino propagation includes interactions, oscillations and the spatial distribution of the source. The numerator includes the model branching fractions. The stated range spans the adopted effective-area/flavor and capture/transport variations.

\section{Solar contrast at comparable high-energy recoil signals}
$\mathrm{BP2}^{\rm c}$ and $\mathrm{BP3}^{\rm c}$ have total capture rates $5.9531\times10^{19}$ and $4.9695\times10^{19}\,\mathrm{s}^{-1}$. Their coherent contributions are $5.9531\times10^{19}$ and $4.9137\times10^{19}\,\mathrm{s}^{-1}$; elastic capture gives only a modest correction for $\mathrm{BP3}^{\rm c}$. The high-energy recoil counts differ by $11.5\%$, rather than forming an exact matched-spectrum pair. At similar transition-normalized capture, the solar source consequently tests information beyond the accessible coupling.

For $\mathrm{BP2}^{\rm c}$, $\Gamma_A=2.9900\times10^7\,\mathrm{s}^{-1}$ and $F_\Gamma\simeq10^{-12}$. For $\mathrm{BP3}^{\rm c}$, the orbital calculation gives $F_\Gamma=0.731$--$0.798$, $\Gamma_A=(1.82$--$1.98)\times10^{19}\,\mathrm{s}^{-1}$ and $\mathcal R_\odot=0.762$--$0.925$. BP1 has $C_{\rm tot}=1.9912\times10^{21}\,\mathrm{s}^{-1}$, $\Gamma_A=1.0624\times10^{17}\,\mathrm{s}^{-1}$ and $F_\Gamma=1.0670\times10^{-4}$ in the reference evaluation. Its low-speed $11$, $12$ and $22$ rates are $1.4725\times10^{-31}$, $2.8262\times10^{-28}$ and $3.9076\times10^{-26}\,\mathrm{cm}^3\,\mathrm{s}^{-1}$; the neutrino source is nevertheless almost entirely ground-state annihilation, because the excited population is depleted.

The dominant neutrino-producing channels remain comparable. $\mathrm{BP2}^{\rm c}$ has branching weights approximately $50.8\%$ into $WW$, $40.9\%$ into $ZZ$, $4.8\%$ into $t\bar t$ and $3.5\%$ into $hZ$; $\mathrm{BP3}^{\rm c}$ has $54.3\%$, $43.5\%$, $0.06\%$ and $2.1\%$, respectively. Ground-state annihilation dominates both sources. Their contrast primarily reflects annihilation efficiency and spatial evolution.

The small recoil splitting is particularly transparent in the signed spectrum of $\mathrm{BP3}^{\rm c}$: $(t_1,t_2,t_3)=(+737.714343,-737.714568,-739.831203)\,\GeV$. The accessible $225\,\keV$ gap compares $|t_1|$ and $|t_2|$, whereas the singlino mixes with the same-sign Higgsino separated by $|t_3|-|t_2|=2.1166\,\GeV$. The corresponding mixing entry is approximately $\Delta_{HS}\sqrt{f}\simeq10.6\,\mathrm{MeV}$, substantially larger than $\delta_{21}$ but small compared with $\Delta_{HS}$. Varying the singlino diagonal mass changes the residual physical degeneracy without imposing $|b|\lesssim\delta_{21}$. $\mathrm{BP2}^{\rm c}$ realizes the same distinction with a singlino ground state. 

\section{Bidirectional SHH/HSH tests}
\label{SM:sec:cross}
We fit six solar-compatible references with physical spectra of the opposite ordering, retaining SI, SD and coherent transitions and the common acceptance. The directional Poisson deviance of the Letter uses $10\,\keV$ bins over $0$--$600\,\keV$; its high-energy restriction uses $200$--$300\,\keV$. We compare accepted true energies and $23\,\keV$ Gaussian broadening. Solar-selected alternatives satisfy $\mathcal R_\odot^{\max}<1$; the solar information enters as a selection, not a neutrino spectral likelihood.

\begin{figure}[t]
\centering\includegraphics[width=240pt,height=186pt]{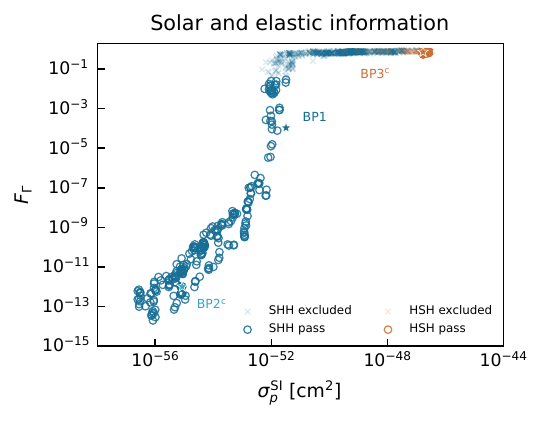}
\caption{Solar--elastic information in SHH/HSH alternatives with explicit solar calculations. Open circles satisfy the adopted solar comparison; pale crosses exceed it. Colors distinguish the orderings. Star symbols mark the benchmark point BP1 and comparison spectra from Table~\ref{tab:bench}. }
\label{SM:fig:crossdiagnostics}
\end{figure}

Figure~\ref{SM:fig:crossdiagnostics} illustrates the complementary solar and elastic information: the tested solar-compatible SHH alternatives have both weak SI scattering and low annihilation efficiency, while HSH retains a stronger SI signal and electroweak annihilation. Their proton SI cross sections span $2.47\times10^{-57}$--$3.20\times10^{-52}$ and $1.67\times10^{-47}$--$2.64\times10^{-47}\,\mathrm{cm}^2$, respectively. Their $F_\Gamma$ ranges are $1.96\times10^{-14}$--$0.0294$ and $0.559$--$0.713$.

The SHH/HSH comparison is not restricted to the $\Delta Q\le11.83$ window used in Fig.~\ref{fig:parameters} of the Letter. Stars in Fig.~\ref{SM:fig:crossdiagnostics} mark BP1, $\mathrm{BP2}^{\rm c}$ and $\mathrm{BP3}^{\rm c}$. The ranges quoted above refer to the comparison sample; BP1 and $\mathrm{BP3}^{\rm c}$ are additional benchmarks not included in that sample. The minimum deviances are obtained over finite samples and do not establish global minima in the continuous parameter space. Shared detector nuisances, backgrounds and scalar cancellations must be included before converting the contrast into an experimental reach.

\section{Elastic cancellations and the HHS alternative}
The SI amplitude is a sum over scalar exchanges,
\begin{equation}
 f_N\propto\sum_a\frac{g_{h_a\chi\chi}g_{h_a NN}}{m_{h_a}^2},
 \qquad\sigma_N^{\rm SI}=\frac4\pi\mu_{\chi N}^2|f_N|^2.
\end{equation}
The normalization of $f_N$ follows the nucleon convention above. Scalar-exchange cancellations can suppress this amplitude independently of $D_1$~\cite{SM:Blind,SM:Roy}. In the heavy-scalar decoupling limit a Higgsino amplitude retains gaugino terms proportional to
\begin{equation}
 [1+\operatorname{sgn}(\mu)\sin2\beta]
 \left[\frac{g_1^2}{M_1-|\mu|}+\frac{g_2^2}{M_2-|\mu|}\right].
\end{equation}
This restricted relation illustrates the sign and $\tan\beta$ dependence; additional Higgs or singlet exchange can modify it. Neither small singlino mixing nor a small transition coupling alone guarantees an SI blind spot.

For a Majorana ground state, $\bar\chi\gamma^\mu\chi=0$, while its diagonal $Z$ interaction contains $D_1\bar\chi\gamma^\mu\gamma^5\chi$. After integrating out the $Z$, the leading quark operator is
\begin{equation}
 \begin{aligned}
 \mathcal L_{\rm SD}&=\sum_qd_q
 (\bar\chi\gamma^\mu\gamma^5\chi)
 (\bar q\gamma_\mu\gamma^5q),\\
 d_q&\propto G_FD_1T_3^q.
\end{aligned}
\end{equation}
Matching onto nucleons gives $a_N\propto\sum_qd_q\Delta q^{(N)}$ and hence $\sigma_N^{\rm SD}\propto D_1^2$. An SI cancellation constrains the scalar sum $f_N$, whereas the SD coupling measures the up/down Higgsino imbalance. They are distinct conditions: suppressing SI scattering need not remove diagonal axial scattering. Heavy squarks make the $Z$ term dominant for BP4.

For BP4, $D_1=-0.0101434$ and $F_{H,1}=0.994973$. Its neutron SI cross section is $7.40\times10^{-49}\,\mathrm{cm}^2$, while the free-proton and free-neutron SD normalizations are $4.19\times10^{-42}$ and $3.21\times10^{-42}\,\mathrm{cm}^2$. The coherent transition strength is large but its $2.98355\,\GeV$ splitting makes it kinematically inaccessible even in the solar interior. The percent-level up/down Higgsino imbalance sustains elastic axial scattering. Retaining the signed nucleon amplitudes is essential to the high-energy xenon prediction.

For elastic BP4, $C\simeq4.14\times10^{19}\,\mathrm{s}^{-1}$ and $\Gamma_A\simeq2.07\times10^{19}\,\mathrm{s}^{-1}$ in the thermal-core approximation, with a number-equilibration time of $5.80\times10^8$ years. Its low-speed annihilation rate is about $4.29\times10^{-26}\,\mathrm{cm}^3\,\mathrm{s}^{-1}$, with $48.2\%$ into $WW$, $38.4\%$ into $ZZ$, $6.3\%$ into $t\bar t$ and $7.0\%$ into $hZ$. The solar range $0.508$--$0.777$ includes the adopted capture and neutrino-yield variants; $hZ$ uses an approximate yield decomposition. Closing the transition makes this efficiently annihilating Higgsino capture limited, conditional on the assumed thermalization.

\section{Target complementarity and composition identification}
For maximum incident speed $v$, the endothermic recoil endpoints are
\begin{equation}
 E_\pm=\frac{\mu_{\chi A}^2v^2}{2m_A}
 \left(1\pm\sqrt{1-\frac{2\delta_{21}}{\mu_{\chi A}v^2}}\right)^2.
\end{equation}
The interval exists for $v\ge\sqrt{2\delta_{21}/\mu_{\chi A}}$, and $\delta_{\max}=\mu_{\chi A}v^2/2$. For BP1, $^{131}$Xe gives $v_\star=765.1\,\mathrm{km\,s}^{-1}$, $E_\star=256.5\,\keV$ and $E_- =142.9\,\keV$ at $798\,\mathrm{km\,s}^{-1}$. On $^{184}$W these become $670.7\,\mathrm{km\,s}^{-1}$, $237.6\,\keV$ and $70.6\,\keV$. The maximal splitting rises from $347.3$ to $451.9\,\keV$. These are isotope-specific kinematic statements, independent of detector acceptance. Natural-isotope weights, tungsten spin response, CaWO$_4$ composition and acceptance are needed for an event-ratio forecast~\cite{SM:Frontier,SM:CRESST}.

The ordering conclusions depend on the halo high-speed tail, nuclear responses, solar abundances, orbital cooling, excited-state lifetime and neutrino propagation. Changing the halo shape alters $K_\odot/K_{\rm Xe}$; rescaling the local density alone does not. Near scalar or axial cancellations, loop corrections may either restore elastic scattering or further suppress it. In the latter case, incomplete cooling leaves smaller spatial overlaps and can weaken the Higgsino solar signal~\cite{SM:Higgsino}. We have not searched for simultaneous loop-induced cooling cancellations. Any such alternative must be propagated through the orbital evolution together with its recoil prediction. The residual pole splitting also needs stability within the adopted perturbative spectrum treatment.

\section{Collider searches and prospects}
\label{SM:sec:collider}
We compare the predicted decay patterns and $13\,\mathrm{TeV}$ production rates with ATLAS and CMS simplified-model constraints using SModelS 3.2~\cite{SM:SModelS}. Electroweakino, slepton and sneutrino production, together with the evaluated strong-production channels, are included. The benchmark spectra retain the full gaugino sector. We compare soft-decay compression thresholds of $1$ and $5\,\GeV$.
The largest predicted-rate/upper-limit ratios across the covered searches and these settings are $r_{\rm LHC}=0.0572$, $0.0141$, $0.0163$ and $0.0401$ for BP1-BP4. For BP1 and BP4 the summed electroweak leading-order rates are $32.7$ and $30.9\,\mathrm{fb}$; the calculated strong-production subsets add $6.04\times10^{-4}$ and $1.57\times10^{-4}\,\mathrm{fb}$. 

Charged coannihilators are essential to this test. In BP1 and BP4, the light stau is $5.68$ and $2.89\,\GeV$ above the neutralino, so the two-body decay to $\tau\chi_1^0$ is open. Their proper decay lengths are $6.88\times10^{-12}$ and $3.17\times10^{-10}\,\mathrm{m}$, removing the heavy stable charged-particle signature targeted by ATLAS-SUSY-2016-32~\cite{SM:ATLASHSCP}. Chargino decays to an on-shell stau are closed; calculated partial widths imply proper-length upper bounds of $65.3$ and $5.68\,\mu\mathrm{m}$. The corresponding lifetimes are well below $0.01\,\mathrm{ns}$, the lower end of the recent ATLAS disappearing-track search~\cite{SM:ATLASTrack}. Its Higgsino mass reach near a nanosecond cannot therefore be applied as a lifetime-independent bound to these spectra. Including coannihilation gives $\Omega_\chi h^2=0.12068$ and $0.11972$, respectively.

Prompt compressed-spectrum searches instead probe initial-state radiation, missing momentum, soft leptons and low-momentum charged decay products~\cite{SM:CMSCompressed,SM:ATLASSoft}. Their sensitivity depends on the chargino and heavier-neutralino gaps, which are independent of the $\mathcal O(100\,\keV)$ recoil transition. For example, the chargino gaps of BP1 and BP4 are $1.36$ and $2.17\,\GeV$. At higher luminosity, improved soft-object reconstruction and initial-state-radiation categories can test this compressed sector, while heavier neutralinos provide additional visible cascade channels~\cite{SM:HLLHC}. Disappearing tracks or displaced vertices become relevant on branches with longer charged-particle lifetimes. These collider observables constrain the coannihilating charged spectrum and complement recoil--solar measurements of the neutral states.

\end{document}